# Non-Markovian dynamics in ice nucleation

Pablo Montero de Hijes

*Faculty of Physics, University of Vienna, A-1090 Vienna, Austria*

Sebastian Falkner

*Institute of Physics, University of Augsburg, 86159 Augsburg, Germany and*
*Faculty of Physics, University of Vienna, A-1090 Vienna, Austria*

Christoph Dellago*

*Faculty of Physics and Research Platform on Accelerating Photoreaction*
*Discovery (ViRAPID), University of Vienna, A-1090 Vienna, Austria*

In simulation studies of crystallisation, the size of the largest crystalline nucleus is often used as a reaction coordinate to monitor the progress of the nucleation process. Here, we investigate, for the case of homogeneous ice nucleation, whether the nucleus size exhibits Markovian dynamics, as assumed in classical nucleation theory. Using 300 independent nucleation trajectories generated by molecular dynamics, we evaluate the mean recurrence time required to reach selected values of the largest nucleus size. Early recurrences consistently take longer than later ones, revealing a clear history dependence and thus non-Markovian dynamics. To identify the slow modes underlying this behaviour, we analyse several structural descriptors of the nucleus, observing subtle but systematic differences between nuclei at early and late recurrences. By training a neural network on 2,700 short trajectories to learn the committor, we identify relevant collective variables. Based on these features, symbolic regression provides a compact approximation of the committor, i.e., an improved reaction coordinate, which we subsequently test for Markovianity.

*Corresponding author:
christoph.dellago@univie.ac.at

## I. INTRODUCTION

Homogeneous ice nucleation is a fundamental process whose microscopic mechanism remains only partially understood and continues to attract significant attention [1–3]. Classical nucleation theory (CNT) [4] is often used to describe this phenomenon, sometimes as part of computational approaches such as Seeding [5–7]. According to this theory, crystallisation proceeds via the formation of an ice nucleus in the supercooled liquid. A nucleation barrier hinders growth, and only nuclei that, by a rare fluctuation, reach the critical size corresponding to this barrier can grow further and complete crystallisation. In its standard formulation, CNT relies on the Markovianity of the dynamics of nucleus size; however, generalisations that include memory effects have been derived [8]. In addition, CNT relies on several simplifying assumptions, such as the sphericity of the nucleus and the presence of a single crystalline phase. Moreover, CNT typically neglects subtleties in the thermodynamics of curved interfaces [9–11]. In crystalline solids, additional complications arise from strained states, defects, composition, or faceting. Although it has been suggested that such complexities may be subsumed into suitably chosen bulk reference states [12–15], their impact on the Markovianity of the process is unclear.

The assumptions made in CNT restrict the description of the nucleation process to a coarse-grained coordinate that omits many relevant slow degrees of freedom. Particularly in ice nucleation, assuming a single crystalline phase neglects the possible stacking of hexagonal (Ih) and cubic (Ic) ice [16–22] or the five-fold twin boundary defect identified by Ref. [23]. In Ref. [24], structural and dynamical heterogeneities of supercooled water were identified as key factors influencing the ice nucleation process. Beyond ice, studies on colloids have shown that prestructured surface particles play an important role in nucleation, an effect that is not captured by the nucleus $n$ alone [25]. For the Lennard–Jones (LJ) system, the dynamics of the largest nucleus size is manifestly non-Markovian [8, 26], and Moroni *et al.* demonstrated that crystallinity, shape, and structural composition (BCC/HCP/FCC) contribute importantly to the reaction coordinate [27]. Together, these studies show that crystallisation cannot, in general, be reduced to one-dimensional kinetics along the largest cluster size alone, and that memory effects arise naturally when essential slow variables are omitted.

Here, using unbiased nucleation trajectories of the mW model obtained by straightforward molecular dynamics, we show that the dynamics projected onto the largest nucleus size $n$ exhibits clear signatures of non-Markovian behaviour. In particular, we show that the average recurrence time needed to reach a certain nucleus size strongly depends on the number of times the particular nucleus size has been reached before, a clear memory effect. This strategy has been used previously to reveal pronounced memory effects in the nucleation dynamics of the LJ-system [26]. We then analyse a broad set of structural descriptors and use a machine-learned committor to identify additional collective variables relevant to the reaction coordinate. Interestingly, an analysis of the dynamics of the approximated committor indicates that it displays signatures of non-Markovian behaviour. The committor is considered to be the optimal reaction coordinate [28, 29], hence our results suggest that even further features should be considered.

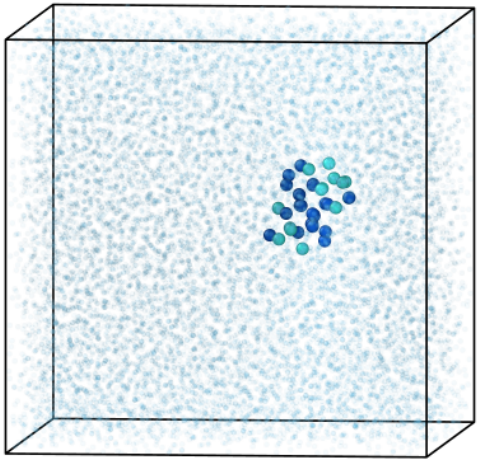

FIG. 1. Snapshot of a crystalline nucleus with Ih-like and Ic-like molecules shown in blue and cyan, respectively, embedded in the supercooled liquid with molecules depicted partially transparently.

The remainder of the article is organised as follows. In Sec. II we describe the simulation setup, the structural analysis of the crystalline nuclei, and the machine-learning workflow used to construct an approximate committor. In Sec. III we first quantify the non-Markovian dynamics of the largest nucleus size (Sec. III A), then analyse structural features associated with this behaviour (Sec. III B), introduce a machine-learned reaction coordinate based on the committor (Sec. III C), and finally compare the Markovianity of the largest nucleus size with that of the committor-based coordinate (Sec. III D). In Sec. IV, we summarise our main findings and discuss their implications for constructing improved reaction coordinates for ice nucleation and related rare-event processes.

## II. METHODS

Molecular dynamics simulations of the coarse-grained mW water model [30] were carried out with LAMMPS [31] using the parametrisation of Ref. [32]. Each system contained $N = 12{,}800$ molecules in a periodic cubic box. Simulations were performed at $T = 222$ K and $p = 1$ bar using a Nosé–Hoover thermostat and barostat. Under these conditions, spontaneous nucleation occurs on accessible timescales of nanoseconds, ensuring that no bias is required and allowing a clean assessment of the Markovianity of the largest nucleus size.

We employed the CHILL+ algorithm [33] via OVITO [34] to classify each molecule as liquid, cubic-ice (Ic), hexagonal-ice (Ih), or others. Clusters of ice-like molecules (Ic plus Ih) were identified using a 3.5 Å cutoff distance. The number of molecules in the largest cluster ($n = n_{Ic} + n_{Ih}$) was then recorded along 300 independent nucleation trajectories. Fig. 1 shows the snapshot of a nucleus that spontaneously formed. For each trajectory, the times $t_n$ at which $n$ crossed selected threshold values from below were determined. A recurrence time, $\tau_n$, was defined as $t_n(i) - t_n(i-1)$, where $i$ indexes successive upward crossings of $n$. Fig. 2 a) shows $n(t)$ for two example trajectories, including an inset clarifying how recurrence times are obtained. Recurrence times were grouped in blocks of four crossings, and the three recurrence intervals within each block were averaged. The block-averaged recurrence times were then averaged over all trajectories. The mean first passage time to reach a certain $n$ is shown in Fig. 2 b).

Structural features for the molecules of the largest nucleus were obtained with OVITO [34] and pyscal [35]. OVITO was used to compute the Ic/Ih composition, the radius of gyration $R_g$, the components of the gyration tensor (used to determine shape metrics such as the asphericity $\mathcal{K}$), and the mean and standard deviation of the Voronoi molecular volume $\mathcal{V}$ and coordination number $\mathcal{C}$. Pyscal was used to compute the tetrahedrality parameter of Uttormark et al. [36], $\theta$, as well as the Lechner-Dellago order parameters $\bar{q}_4$ and $\bar{q}_6$ [37]. Molecular features were averaged among the nucleus molecules. Several additional features were derived from these quantities to facilitate the training of the neural network.

The training of the committor neural network follows the workflow described by Jung et al. [38]. We employ a multilayer perceptron (see SI for network architecture) and train on $N = 2{,}700$ fleeting trajectories, each labelled $y_i = -1$ if state A (liquid, $n < 10$) or $y_i = 1$ if state B (solid, $n > 60$) was reached. The initial configurations for these trajectories are obtained by uniformly sampling configurations where $n$ is between 10 and 60 molecules from the original 300 trajectories in order to then reinitialize the velocities. We define the predicted committor as $p_B(\mathbf{x}, \mathbf{w}) = 1/(1 + \exp(-q(\mathbf{x}, \mathbf{w})))$ with $\mathbf{x}$ being the normalized input collective variables and $\mathbf{w}$ being the network weights. The loss function for training the network is accordingly $L(\mathbf{w}|\theta) = \sum_{i=0}^{N} \log[1 + \exp(y_i q(\mathbf{x}_i, \mathbf{w}))]$. We train a committee of 10 neural networks with a batch size of 16 samples each for 100 epochs using a cosine annealing learning rate schedule between $10^{-2}$ and $10^{-6}$. We estimate attribution values as in Jung et al. [38] by permutation of the input features, followed by symbolic regression using dCGP [39], keeping the four most important features. In contrast to previous studies, we perform the regression with respect to the network output instead of the shooting data, as our initial dataset only includes one shot per point. The regression is run 8 times for 300 generations, allowing for summation, subtraction, multiplication, division, square and cube operators. We then select the best fitting expression that reasonably extrapolates for values outside of the training range (e.g., large cluster sizes should yield high committor probabilities).

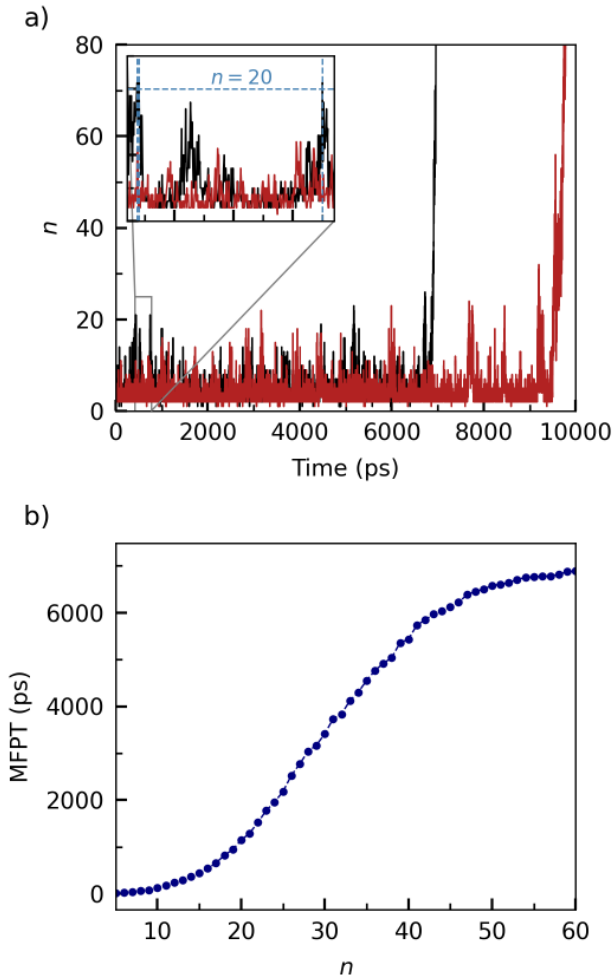

FIG. 2. a) Time evolution of the largest cluster size $n$ for two example trajectories, shown in black and red. The inset zooms in on one part of the trajectory where, if we define the threshold to be $n = 20$ (horizontal line), then we find three crossing times $t'_{20}$ (vertical lines), hence, defining two recurrence times (the time difference between the second and first, and the third and second crossings). b) Mean first passage time (MFPT) with respect to $n$.

## III. RESULTS

### A. Non-Markovian dynamics of the nucleus size

A Markovian description of nucleation, as postulated in CNT, assumes that the future evolution of the largest cluster size $n$ depends only on its current value. To test this assumption, we analyse the statistics of upward crossings of selected thresholds of $n$. Fig. 3 shows the mean recurrence times $\langle \tau_n \rangle$ for blocks of four consecutive crossings. If the dynamics projected onto $n$ were Markovian, all blocks at a given $n$ would yield identical recurrence times. Instead, the first block exhibits systematically longer recurrence times—up to nearly twice those of later blocks—demonstrating that trajectories reaching the same $n$ retain memory of how they reached that value. This block dependence constitutes a clear signature of non-Markovian behaviour.

Further evidence for non-Markovianity is provided in Fig. 4, which reports the average minimum and maximum values of the nucleus size explored between consecutive crossings of $n$. The first block clearly explores systematically smaller nuclei, both in their minima and maxima, compared to later blocks. The strong dependence of these excursion ranges on the crossing index indicates that the system retains some memory of previous states, at least on the initial stage of nucleation.

Together, these observations demonstrate unambiguously that the dynamics of the largest nucleus size, $n$, is intrinsically non-Markovian. Consequently, $n$ alone cannot serve as a complete reaction coordinate for nucleation. Naturally, some arbitrariness is inherent in the definition of $n$, and one could argue that our specific criterion for identifying the largest nucleus might itself be responsible for the observed non-Markovian behaviour—that is, a different definition of $n$ might yield a coordinate with more Markovian dynamics. We acknowledge this possibility, yet we emphasize that our definition follows standard practices in the ice-nucleation community [33, 40–47]. In studies where cubic ice is not considered, the largest cluster is identified using criteria based solely on $\bar{q}_6$ [7, 48, 49] and alternative definitions of $n$ have also been used [20, 24]. A systematic comparison of the dynamics of $n$ for different definitions is an interesting subject for future work.

### B. Structural features

What the largest nucleus size misses to be Markovian may lie in subtle structural relaxations. To probe this hypothesis, we computed a series of structural descriptors for the molecules belonging to the largest ice nucleus at every timestep of each trajectory. These descriptors were chosen to capture local structural motifs that may be overlooked when using only the largest nucleus size.

To account for the crystalline composition of the nucleus, we computed the ratio $n_{Ih(c)}/n$, where $n_{Ih(c)}$ is the number of hexagonal(cubic)-ice-like molecules within the largest nucleus. To characterise tetrahedral arrangements, we evaluated the Uttormark tetrahedral order parameter $\theta$ for each molecule of the largest nucleus and averaged it over them, obtaining $\Theta$. In this convention, lower values of $\theta$ (and hence of $\Theta$) correspond to more tetrahedral local environments. To quantify local crystallinity, we computed the local order parameters $\bar{q}_4$ and $\bar{q}_6$ and averaged them over the molecules in the largest nucleus, yielding $\bar{Q}_4$ and $\bar{Q}_6$, respectively.

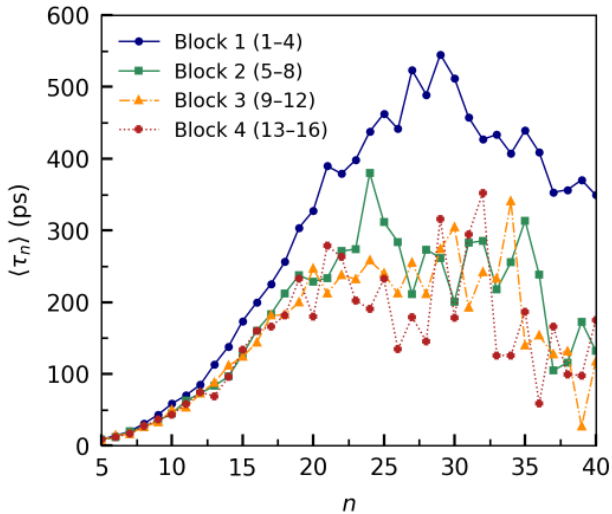

FIG. 3. Mean recurrence times $\langle \tau_n \rangle$ as a function of the largest nucleus size threshold, averaged over blocks of four consecutive crossings. The first block (solid line + circles in blue) reaches about twice as long recurrence times, revealing non-Markovian behaviour.

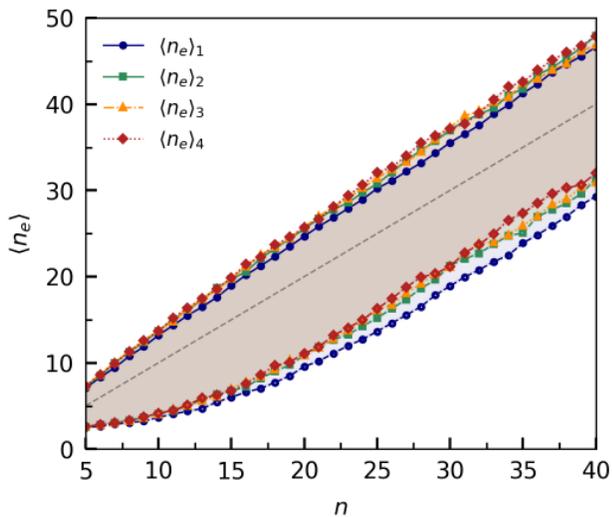

FIG. 4. Average maximum (solid) and minimum (dashed) values of $n$ visited during excursions between threshold crossings per block, $\langle n_e \rangle$. The first block, $\langle n_e \rangle_1$, shows both maximum and minimum values that are systematically lower than later blocks, remarking the non-Markovian dynamics of $n$.

To characterise the overall geometry of the nucleus, we first computed its radius of gyration $R_g$. We then assessed the shape of the nucleus using the eigenvalues of the gyration tensor, denoted by $\lambda_1 \leq \lambda_2 \leq \lambda_3$. From these, we obtained the anisotropy $\mathcal{I}$,

$$\mathcal{I} = \frac{3}{2} \left( \frac{\lambda_1^2 + \lambda_2^2 + \lambda_3^2}{(\lambda_1 + \lambda_2 + \lambda_3)^2} - \frac{1}{3} \right), \quad (1)$$

the acylindricity $\mathcal{J}$,

$$\mathcal{J} = \lambda_2 - \lambda_1, \quad (2)$$

and the asphericity $\mathcal{K}$,

$$\mathcal{K} = \lambda_3 - \frac{1}{2}(\lambda_1 + \lambda_2). \quad (3)$$

We also attempted to detect additional information on density inhomogeneities within the nucleus through the mean and standard deviation of the Voronoi molecular volumes, $\mathcal{V}_m$ and $\mathcal{V}_\sigma$, respectively. Analogously, for the Voronoi-based coordination number of the molecules in the nucleus, we defined $\mathcal{C}_m$ and $\mathcal{C}_\sigma$ as the mean and standard deviation of the coordination number distribution.

Finally, we combined these primary descriptors into a set of derived features that emphasise different physical aspects of the nucleus and might ease the training of the network. Specifically, we defined

$$\mathcal{E}_1 = n \cdot \bar{Q}_6, \quad (4)$$

$$\mathcal{E}_2 = \frac{\mathcal{K}}{R_g^2}, \quad (5)$$

$$\mathcal{E}_3 = \frac{\mathcal{V}_\sigma}{\mathcal{V}_m}, \quad (6)$$

$$\mathcal{E}_4 = \frac{\mathcal{C}_\sigma}{\mathcal{C}_m}, \quad (7)$$

$$\mathcal{E}_5 = \frac{1 + \mathcal{E}_3 + \mathcal{E}_4 - \bar{Q}_6}{3}, \quad (8)$$

$$\mathcal{E}_6 = \frac{n}{R_g^3}. \quad (9)$$

Here, $\mathcal{E}_1$ measures an "ordered" nucleus size, $\mathcal{E}_2$ combines shape anisotropy with the nucleus compactness, $\mathcal{E}_3$ and $\mathcal{E}_4$ quantify relative fluctuations in Voronoi volume and coordination, $\mathcal{E}_5$ mixes density heterogeneity and local order into a single descriptor, and $\mathcal{E}_6$ is a measure of the effective number density of the nucleus.

Similar to the procedure used to compute $\langle \tau_n \rangle$, we evaluate the average value of each structural descriptor within each occurrence block for a given threshold $n$. This allows us to determine whether a particular structural feature changes between early and late occurrences of the same largest nucleus size. From this analysis, we see that several descriptors do not differ among blocks, but $\bar{Q}_6$, $\bar{Q}_4$, $\Theta$, and $\mathcal{C}_m$ exhibit small but systematic differences between occurrence blocks, as shown in Fig. 5. Early occurrences of a given nucleus size tend to display slightly lower orientational order and reduced tetrahedral arrangement compared with later ones. We recall that a higher $\Theta$ means less tetrahedral arrangement. The coordination also increases modestly across blocks for the larger nuclei. These trends may suggest that the internal structure of clusters of a given size evolves during the trajectory and retains memory of the history of the system, consistent with the recurrence-time analysis.

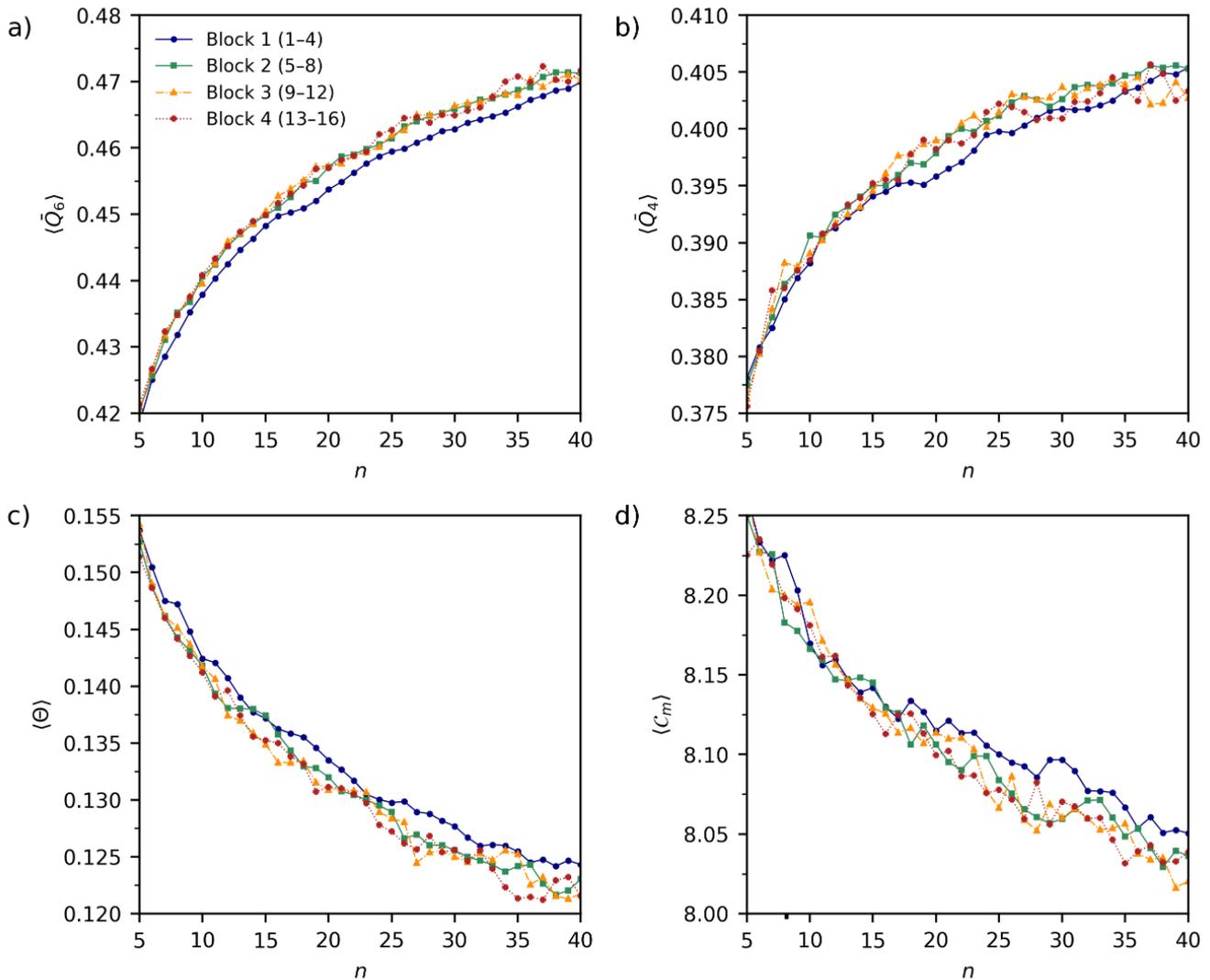

FIG. 5. a) $\langle \bar{Q}_6 \rangle$, b) $\langle \bar{Q}_4 \rangle$, c) $\langle \Theta \rangle$, and d) $\langle \mathcal{C}_m \rangle$ for the largest nucleus, averaged per occurrence block for each threshold value. The legend in panel a) applies to all panels.

It is worth stressing, however, that the interpretation of individual descriptors in this block analysis must be approached with caution. The differences observed between blocks do not, by themselves, determine how a given descriptor affects the nucleation probability, because its influence may be outweighed or even counterbalanced by other, more relevant variables. In particular, nuclei appearing in the first block could, in principle, carry a structural signature that would favour nucleation, yet at the same time exhibit other characteristics that have a much stronger and opposite impact on the nucleation probability.

Although the variations in any single feature are small, taken together, they suggest that nucleation proceeds along additional slow directions of the configuration space that are not captured by the largest nucleus size alone. Identifying if any of these descriptors plays a relevant role requires further analysis. In the next subsection, we address this by performing a committor analysis based on extensive sets of short trajectories, followed by machine-learning classification and symbolic regression to determine which structural features are essential components of the reaction coordinate.

### C. Machine learned reaction coordinate

The previous analyses have shown that the largest nucleus size alone is not sufficient to predict the future behaviour of the system. Therefore, we aim to find a better reaction coordinate to describe nucleation that minimises non-Markovian behaviour. In general, the ideal reaction coordinate for a rare event such as nucleation is the committor $p_B(\mathbf{x})$ [28, 29], which describes the probability that a trajectory started from a certain configuration $\mathbf{x}$ transitions to the ice phase. While the ground truth committor as a function of all Cartesian coordinates is analytically intractable and can only be obtained by resource intensive simulations, Jung et al. [38] have shown that a neural network can approximate the committor as a function of a set of relevant collective variables. Leveraging these findings, we train a neural network on 2,700 fleeting trajectories to predict the committor for ice nucleation to find additional important degrees of freedom together with their contribution to the nucleation prob-ability.

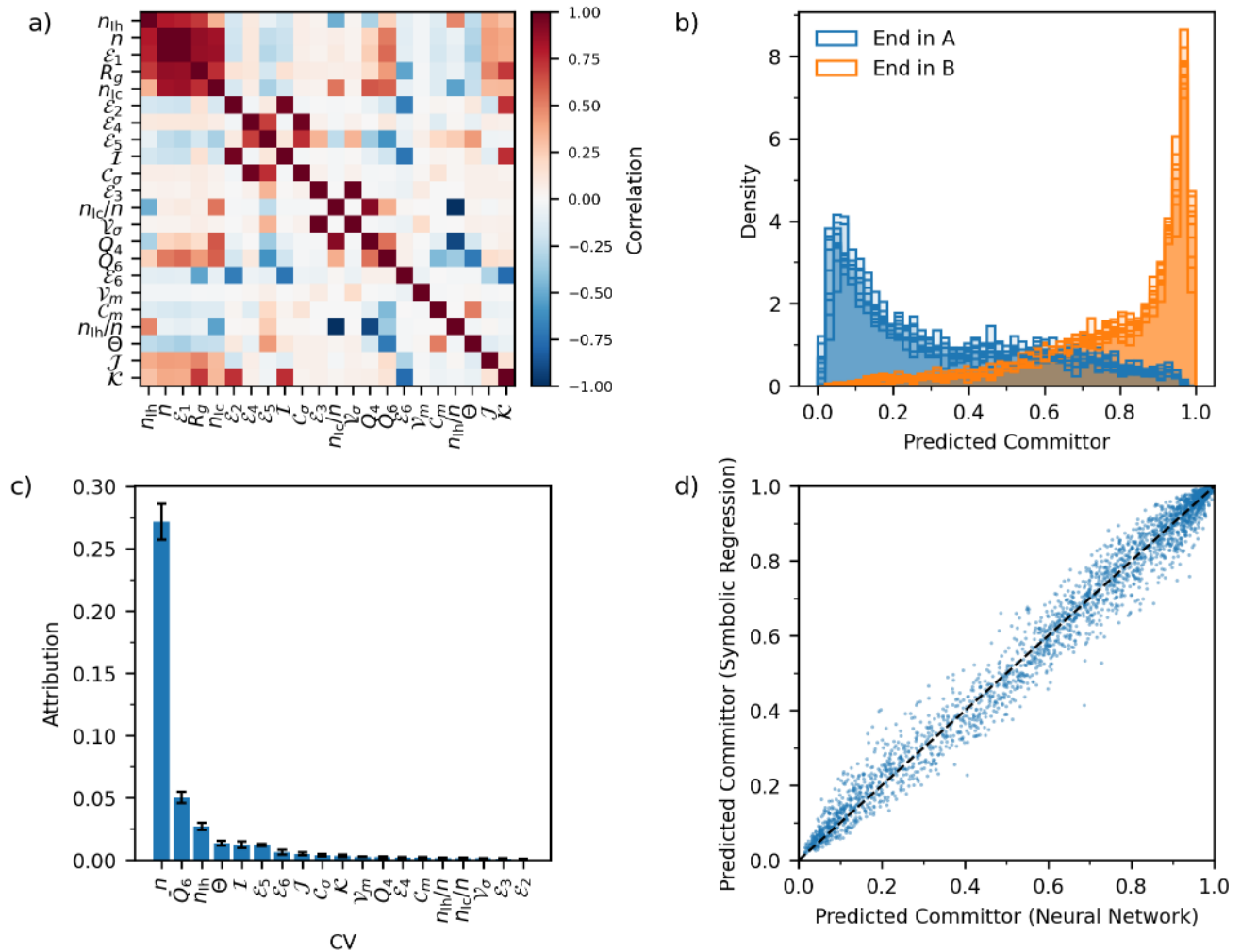

FIG. 6. a) Correlation matrix for a selection of collective variables that describe the nucleus geometry. b) Distribution of network-predicted committor values for trajectories that ended in the liquid (State A) or solid state (State B). The histograms for the ten neural networks in the committee are depicted on top of each other. c) Network attribution values for each input collective variable obtained by permuting each feature separately and observing its impact on the loss function. Higher values indicate important features for the prediction of the committor. d) Comparison of the neural network predicted committor to the committor obtained from symbolic regression.

We first perform a correlation analysis on our broad set of collective variables by calculating their cross-correlations (Fig. 6a). Based on these results, we removed variables that strongly correlated to the nucleus size, namely $\mathcal{E}_1$, $n_{Ic}$ and $R_g$, as they interfere with the later attribution analysis. After training the neural network, we see that it can successfully distinguish configurations from which the system tends to transition to ice or fall back to the liquid state (Fig. 6b). Each input feature of the network is then assigned an attribution corresponding to its importance to the predicted committor (Fig. 6c). We observe that, although the nucleus size remains the most important parameter to predict the committor probability by a clear margin, multiple other

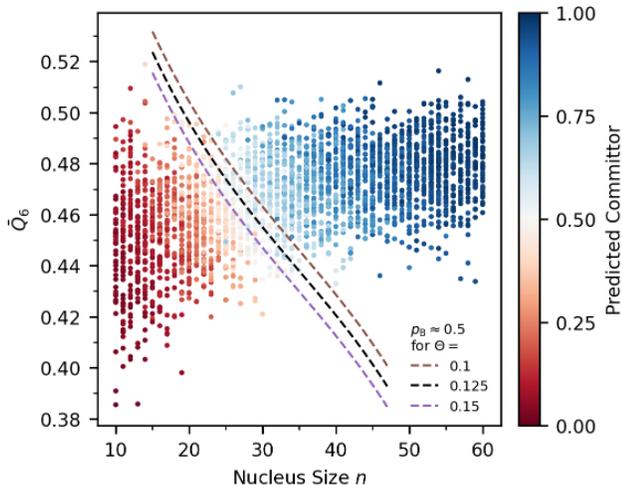

FIG. 7. Committor landscape as a function of the nucleus size $n$, the nucleus-averaged Lechner-Dellago order parameter $\bar{Q}_6$, and the nucleus-averaged tetrahedrality parameter $\Theta$. Each point represents the initial point of a fleeting trajectory in the dataset and is coloured according to the network-predicted committor. The dashed lines represent the $p_B \approx 0.5$ isocommittor lines for different tetrahedralities $\Theta$.

collective variables contribute to the network prediction.

Based on these findings, we perform symbolic regression to find a human-readable expression for the nucleation committor probability. We include as inputs the four most important features: the nucleus size $n$, the order parameter averaged over the nucleus molecules $\bar{Q}_6$, the number of hexagonal ice molecules $n_{1h}$, and tetrahedrality parameter averaged over the nucleus molecules $\Theta$. Using these features, the best fitting expression that also extrapolates well to larger nuclei is:

$$p_B(\tilde{n}, \tilde{Q}_6, \tilde{\Theta}) = \text{sig}\big(0.216\tilde{n}^3 + 1.421\tilde{n} + 0.518\,\tilde{Q}_6 \\ + 0.193\,\tilde{\Theta} + 0.803\big),$$ (10)

where $\text{sig}(x) = 1/[1 + \exp(-x)]$ and $\tilde{n}$, $\tilde{Q}_6$, $\tilde{\Theta}$ are the normalized input features:

$$\tilde{n} = \frac{n - 34.624}{14.722}$$ (11)

$$\tilde{Q}_6 = \frac{\bar{Q}_6 - 0.467}{0.018}$$ (12)

$$\tilde{\Theta} = \frac{\Theta - 0.125}{0.021}$$ (13)

Notably, $n_{1h}$ with the third highest network attribution is not present in the fitted expression, most likely due to its correlation with the largest nucleus size. Still, the committor obtained via the simple analytical expression matches the network predicted committor well (Fig. 6d).

Looking at the committor landscape as a function of these three parameters (Fig. 7), we see that an increasing nucleus size is still the best predictor for a successful phase transition. However, especially for critical nuclei with a committor of approximately 0.5, we also observe that the crystallinity of the nucleus, captured by the $\bar{Q}_6$ parameter, is essential for predicting the committor of the system. This is apparent in Fig. 7, where the $p_B \approx 0.5$ isocommittor line is tilted, indicating that for the same nucleus size (e.g., 27 molecules), the $\bar{Q}_6$ parameter can decide whether the nucleus tends to grow or vanish. Additionally, the distilled committor expression indicates that a high $\Theta$, corresponding to a low tetrahedrality, generally increases the commitment probability to the ice state. Interestingly, the block analysis indicated that early crossings with increased recurrence times are correlated with a low tetrahedrality. However, at the same time, these early crossings are associated with a low $\bar{Q}_6$ which possibly negates the nucleation-promoting effect of a low tetrahedrality.

### D. Comparison of Markovianity between $n$ and $p_B$

We further illustrate the non-Markovian behaviour of the largest nucleus size $n$ by examining the block-averaged values of $p_B$ at upward crossings of $n$. If $n$ were a Markovian reaction coordinate, all crossings of a given $n$ would have the same average committor. Instead, as shown in Fig. 8 a), configurations belonging to the first occurrence block systematically exhibit lower values of $p_B$ than those from later blocks. In other words, nuclei with the same size $n$ are less likely to keep growing when they are encountered early in the trajectory. This dependence of the growth probability on the trajectory history confirms once again that $n$ alone does not provide a Markovian description of the dynamics.

Having further established the non-Markovian dynamics of $n$, we now examine whether the approximate committor $p_B$ offers a more Markovian representation. Since $p_B$ incorporates additional structural information and correlates strongly with the probability that a configuration will proceed to the ice basin, one may expect it to reduce the memory effects observed when projecting the dynamics onto $n$. Fig. 8 b) reports the mean recurrence time $\langle \tau_{p_B} \rangle$ required to return to a given threshold of $p_B$. As in the case of $n$, the first block of crossings (1–4) displays longer recurrence times than later blocks, indicating that memory effects persist. This is not surprising: the additional structural descriptors used to construct $p_B$ are unlikely to span all remaining slow degrees of freedom, and projecting even a improved reaction coordinate may not be perfectly Markovian.

The comparison is made explicit in Fig. 8 c), which shows the block-to-block difference $\Delta\langle \tau_{RC} \rangle_{1-2} = \langle \tau \rangle_{1-4} - \langle \tau \rangle_{5-8}$ for the normalized largest nucleus size and $p_B$ both from the neural network and from the symbolic regression. For $n$, we use the normalization $(n - n_A)/(n_B - n_A)$. Across most of the reaction pathway, $\Delta\langle \tau_{RC} \rangle_{1-2}$ is only slightly smaller for the $p_B$ estimations than for $n$, confirming that $p_B$ exhibits modestly improved Markovian behaviour. The fact that the difference remains nonzero over the entire range shows that the committor-based mapping does not eliminate memory effects.

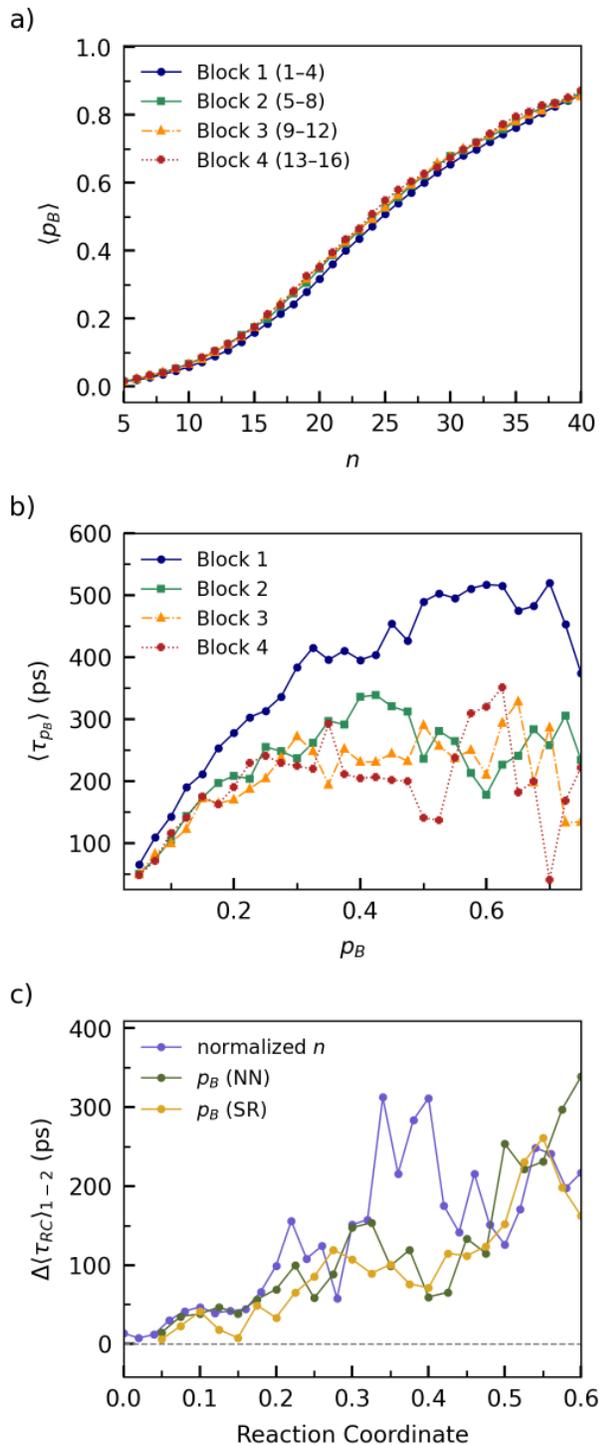

FIG. 8. a) Recurrence-block-averaged committor $\langle p_B \rangle$ as a function of the largest nucleus size threshold, evaluated over successive blocks of four upward crossings. b) Mean recurrence time $\langle \tau_{p_B} \rangle$ return to a given value of $p_B$, again averaged over blocks of four crossings. c) Difference between the first and second blocks of recurrence times, $\Delta \langle \tau_{RC} \rangle_{1-2}$, for three reaction coordinates ($RC$); the normalised cluster size $(n - n_A)/(n_B - n_A)$ and both $p_B$ from the neural network (NN) and symbolic regression (SR). The $\Delta \langle \tau_{RC} \rangle_{1-2}$ of $p_B$ from symbolic regression agrees well with that from the neural network.

Overall, these results demonstrate that $p_B$ provides a modest improvement over $n$ in terms of Markovianity, yet it still does not constitute a fully memoryless reaction coordinate. Additional slow degrees of freedom—not captured by either $n$ or by the structural descriptors used to build $p_B$—continue to influence the dynamics. Nevertheless, the workflow established here provides a robust foundation for future efforts aimed at refining reaction coordinates for ice nucleation by explicitly assessing the Markovianity of their projected dynamics.

## IV. CONCLUSIONS

In this work, we have examined whether the size of the largest ice nucleus, $n$, provides a Markovian description of homogeneous ice nucleation in supercooled water. Using 300 unbiased trajectories for mW-water, we showed that the dynamics projected onto $n$ exhibit clear and systematic history dependence: early crossings of a given $n$ require significantly longer recurrence times than later ones. This behaviour provides direct evidence that the nucleus size $n$ alone is not a Markovian reaction coordinate for ice nucleation. Further support comes from analysing the average minimum and maximum values of $n$ explored between consecutive upward crossings: during the early occurrences, systematically smaller nuclei are visited, again reflecting history dependence.

We probed the origin of these memory effects by analysing a broad set of structural descriptors of the nucleus. Although the differences are modest in magnitude, several features—including local crystalline order, tetrahedrality, and coordination—display systematic trends between early and late occurrences, indicating that structural relaxation proceeds along additional slow degrees of freedom that are not captured by the cluster size alone.

To identify which of these additional variables are most relevant, we trained a neural network to learn the committor from 2,700 short trajectories. An attribution analysis based on this neural network showed that, besides the nucleus size, structural descriptors such as $\bar{Q}_6$ and tetrahedrality play a significant role in determining the nucleation probability. Using this information, symbolic regression yielded a compact approximate analytical expression for the committor $p_B$, providing an interpretable and testable improved reaction coordinate.

Comparing the Markovianity of $n$ and $p_B$ shows that the committor-based coordinate is modestly more Markovian, although memory effects persist. The residual non-Markovianity suggests that the network still misses additional slow modes that influence the dynamics. The reason for this could simply be insufficient training data or possibly an incomplete set of relevant input features. For instance, parameters that characterize the surrounding liquid phase, which are neglected here, could play a role in the nucleation process.

Overall, our study establishes a quantitative framework for assessing the Markovianity of proposed reaction coordinates for ice nucleation. The workflow introduced here, which can also be applied to trajectories obtained with Transition Path Sampling [50] or other trajectory-based rare-event methods, provides the basis for constructing improved low-dimensional coordinates. The approach can easily be extended to more realistic water models with richer structural descriptions, as well as to entirely different nucleation processes.

## ACKNOWLEDGMENTS

This work is dedicated to Carlos Vega, with deep appreciation for the inspiration he brings as a mentor, scientist, and colleague. This research was funded in part by the Austrian Science Fund (FWF) through the SFB TACO 10.55776/F81. For open access purposes, the author has applied a CC BY public copyright license to any author-accepted manuscript version arising from this submission. Computer resources and technical assistance were provided by the Vienna Scientific Cluster (VSC).

## AUTHOR DECLARATIONS

### Conflict of Interest

The authors have no conflicts to disclose

## DATA AVAILABILITY

The data that support the findings of this paper are available upon request.

pp. 1964–1977, 02 1998.